\documentclass{article}
\usepackage{times}
\usepackage{natbib}
\usepackage{astsym}
\begin{document}
\thispagestyle{empty}

\centerline{\Large Skywalking GEMS and UDF}
\vspace{5mm}
\centerline{K.~Jahnke$^1$, S.~F.~S\'anchez$^1$, B.~H\"au\ss ler$^2$}
\centerline{$^1$Astrophysikalisches Institut Potsdam}
\centerline{$^2$Max-Planck Institut f\"ur Astronomie, Heidelberg}
\vspace{5mm}

Recently large high resolution space based imaging surveys for
galaxies have been conducted with the Hubble Space Telescope and its
Advanced Camera for Surveys (ACS). Very prominent are the Hubble Ultra
Deep Field \citep[UDF,][]{udf04}, the observations in the GOODS field
\citep{giav03} and our own GEMS survey \citep[`Galaxy Evolution from
Morphologies and SEDs',][]{rix04}, ordered by increasing sky
coverage. While the UDF covers one ACS pointing or about
$200\arcsec\times200\arcsec$, GEMS covers 70 times as much, nearly
$28\arcmin\times28\arcmin$. With the observed size and depth of the
observations -- the UDF is 1--1.5~mag deeper than the Hubble Deep
Field -- these images are very rich in galaxies of all sizes, shapes,
and -- all mentioned surveys obtained multi-band images -- also
colours. The UDF contains almost 10\,000 galaxies, GEMS about
40\,000.

Both the UDF and GEMS teams created ``true colour'' images from their
data, largely for outreach activity. HST plus its ACS camera have a
number of advantages over other instruments, with one combination
making it fully unique: the large field-of-view and a 0\farcs05
spatial sampling. While this is wonderful for science it poses one
problem for outreach applications: How to view these images? The UDF
in a computer screen sized resolution shows a number of coloured dots
and hides most of the beauty of the very deep space, the GEMS colour
mosaic viewed at a resolution of 1000 pixel in width is largely
black. We think that the GEMS colour mosaic would be best viewed in a
printout about 10~meters on a side (still $\sim$6 pixel per mm!), at
50~cm distance to see all details, which might be difficult to realise
for a widespread audience. While the underlying amount of image data
in JPEG compression is comparably small compared to the science grade
images, it is still 10~Mb for UDF and 175~Mb in case of GEMS, thus not
very friendly for download without access to institute quality internet.

To nevertheless allow people to enjoy the colour mosaics, we put
together a JavaScript web application, that allows to pan around in
the 1:1 UDF and 2:1 binned GEMS images via the WWW, without
downloading all of the images at once, but only the part viewed. In
this way access even with analog modems is possible. We dubbed these
applications ``GEMS Skywalker'' and ``UDF Skywalker'' (names not
sponsored by Lucasfilm). They are available online for free use at\\

\indent
\hspace{1cm}{\tt http://www.aip.de/groups/galaxies/sw/gems/}\hspace{5mm} (GEMS Skywalker)\\
\indent
\hspace{1cm}{\tt http://www.aip.de/groups/galaxies/sw/udf/}\hspace{5mm} (UDF
Skywalker).\\

\noindent
They should work with most Netscape, Internet Explorer and Opera
Versions.
\vspace{3mm}

\noindent
For the realisation of the Skywalkers, we would like to thank both the
UDF and GEMS teams for their work that generated these images. Also a
big Thank You goes to the developers of the DynAPI JavaScript library
\citep{dynapi} that made programming easy.

\bibliography{knuds}
\bibliographystyle{apj}

\end{document}